\documentclass{jfm}
\usepackage{newcent}
\usepackage{verbatim,framed}
\usepackage{bm}
\usepackage{amsmath}
\usepackage{graphicx}
\usepackage{amssymb}
\usepackage[numbers]{natbib}

\makeatletter

\def\clap#1{\hbox to 0pt{\hss#1\hss}}

\def\mathclap{\mathpalette\mathclapinternal}
\def\mathrlap{\mathpalette\mathrlapinternal}
 
\def\mathclapinternal#1#2{%
\clap{$\mathsurround=0pt#1{#2}$}} 
\def\mathrlapinternal#1#2{%
\rlap{$\mathsurround=0pt#1{#2}$}} 
\input  ifpdf.sty
\ifpdf
\usepackage{pdfsync}
\fi
\usepackage{bm}

\makeatother

\newcommand{\rem}[1]{} 

\usepackage[british]{babel}

\begin{document}

\title{The Square Root Depth Wave Equations}

\author[C. J. Cotter , D. D. Holm and J. R. Percival]
{C\ls O\ls L\ls I\ls N\ns  J.\ns C\ls O\ls T\ls T\ls E\ls R$^{1}$,\ns  D\ls A\ls R\ls R\ls Y\ls L\ns D.\ns H\ls O\ls L\ls M$^{2}$\ns \\  \and J\ls A\ls M\ls E\ls S\ns R.\ns P\ls E\ls R\ls C\ls I\ls V\ls A\ls L$^{2}$}
\affiliation{$^{1}$Department of Aeronautics, Imperial College London, SW7 2AZ,
UK \\[\affilskip] $^{2}$Mathematics Department, Imperial College
London, SW7 2AZ, UK}
\date{04/12/09}

\maketitle
\begin{abstract}
We introduce a set of coupled equations for multilayer water waves
that removes the ill-posedness of the multilayer Green-Naghdi (MGN)
equations in the presence of shear. The new well-posed equations 
are Hamiltonian and in the absence of imposed background shear they retain
the same travelling wave solutions as MGN. We call the new model the
Square Root Depth ($\sqrt{D}$) equations, from the modified form
of their kinetic energy of vertical motion. Our numerical results
show how the $\sqrt{D}$ equations model the effects of multilayer
wave propagation and interaction, with and without shear. 
\end{abstract}

\section{Introduction}

The propagation and interactions of internal gravity waves on the
ocean thermocline may be observed in many areas of strong tidal flow,
including the Gibraltar Strait and the Luzon Strait. These waves are
strongly nonlinear and may even be seen from the Space Shuttle \citep{LCHL98},
with their crests moving in great arcs hundreds of kilometres in length
and traversing sea basins thousands of kilometres across. The MGN
model --- the multilayer extension of the well known Green-Naghdi equations
\citep{GN76.,CC99} %
{} --- has been used with some success to model the short term behaviour
of these waves \citep{JoChoi2002}. Nevertheless, MGN and its rigid-lid
version, the Choi-Camassa equation \citeyearpar[hereafter CC]{CC96JFM},
have both been shown by \citet{LB1997} to be \emph{ill-posed} in
the presence of background shear. That is, background shear causes
the linear growth rate of a perturbation to increase without bound
as a function of wave number. Needless to say, this ill-posedness
has made numerical modelling of MGN problematic. In particular, ill-posedness
prevents convergence of the numerical solution, since the energy cascades
to smaller scales and builds up at the highest resolved wave number.
Grid refinement only makes the problem worse. Regularization by keeping
higher-order expansion terms is possible \citep{barros2009isi}, but
such methods tend to destroy the Hamiltonian property of the system
and thus may degrade its travelling wave structure. Nevertheless,
if one is to consider the wave generation problem, one must consider
the effects of both topography and shear.

Thus, the MGN equations must be modified to make them well-posed.
We shall require that the new system: 
\begin{enumerate}
\item is both linearly well-posed and Hamiltonian; 
\item preserves the MGN linear dispersion relation for fluid at rest; 
\item has the same travelling wave solutions as MGN in the absence of imposed
background shear. 
\end{enumerate}
The $\sqrt{D}$ equation we introduce here satisfies these requirements.
The remainder of the paper is organised as follows. In Section 2 we
derive the $\sqrt{D}$ model in the same Euler-Poincar\'{e} variational
framework as for MGN \citep{PCH2008}. In Section 3 we compare the
linear dispersion analysis of $\sqrt{D}$ and MGN, and thus show that
linear ill-posedness has been removed. In Section 4 we show that the
$\sqrt{D}$ and the shear free MGN equations possess the same travelling
wave solutions. In Section 5 we present numerical results that compare
the wave propagation and interaction properties of the $\sqrt{D}$
and GN solutions for a single layer. Finally, in Section 6 we present
numerical results for the $\sqrt{D}$ equations that model the effects
of two-layer wave propagation and interaction, both with and without
a background shear.

\section{The $\sqrt{D}$ Governing Equations}

In this section we derive the $\sqrt{D}$ equations by approximating
the kinetic energy of vertical motion in Hamilton's principle for
a multilayer ideal fluid. Such a system consists of $N$ homogeneous
fluid layers of densities $\rho_{i}$, $i\in[1,\ldots N]$, where
$i=1$ is the top layer and $i=N$ is the bottom layer. Thus, for
stable stratification, $\rho_{i+1}>\rho_{i}$. The $i$-th layer has
a \emph{horizontal} velocity, $\bm{u}_{i}$ and thickness, $D_{i}$; the
interface between the $i$-th and $i-1$-th layer is at depth $h_{i}=-b+\sum_{j=i}^{N}D_{i}$,
for a prescribed bathymetry, $b(x,y)$. We assume columnar motion within each layer (horizontal velocity independent of vertical coordinate, $z$). Incompressibility then implies that vertical velocity is linear in $z$.
Under this ansatz and after a vertical integration, the Lagrangian
for Euler's fluid equations with a free surface becomes, \begin{equation}
\ell=\sum_{i=1}^{N}\frac{\rho_{i}}{2}\int\left[D_{i}|\bm{u_{i}}|^{2}
+\frac{D_{i}}{3}
\left(
w_{i}^{2}\Big|_{z=\mathrlap{h_{i+1}}}
\, + \,
w_{i}\Big|_{z=\mathrlap{h_{i}}} \; w_{i}\Big|_{z=\mathrlap{h_{i+1}}}
\; + \,
w_{i}^{2}\Big|_{z=h_{i}}
\right)
-g[h_{i}^{2}-h_{i+1}^{2}]\right]
\mathrm{d}x\, \mathrm{d}y.
\label{eq:MGN_lagrangian}\end{equation}
 In this Lagrangian, the $w_{i}$ should be viewed as functions of
layer velocities, layer thicknesses and their partial derivatives.
The fluid  momentum equations are obtained from Euler-Poincar\'{e} theory
\citep{HMR1998} as\[
\frac{\partial}{\partial t}\frac{\delta\ell}{\delta\bm{u}_{i}}
+
\nabla\cdot\left[\bm{u}_{i}\frac{\delta\ell}{\delta\bm{u}_{i}}\right]
+
\nabla\bm{u}_{i}^{T}\cdot\frac{\delta\ell}{\delta\bm{u}_{i}}
=
D_{i}\nabla\frac{\delta\ell}{\delta D_{i}}.\]
 The system is closed by the layer continuity equations,\begin{equation}
\frac{\partial D_{i}}{\partial t}+\nabla\cdot D_{i}\bm{u}_{i}=0.\label{eq:D_continuity}\end{equation}
 To obtain the MGN equations one sets $w_{i}$ in (\ref{eq:MGN_lagrangian})
equal to the vertical component of the fluid velocity, represented
in terms of the material time derivative as 
\begin{align}
w_{i} & 
=(h_{i}-z)\nabla\cdot\bm{u}_{i}+\bm{u}_{i}\cdot\nabla h_{i}-\sum_{\mathclap{j=i}}^{N}\nabla\cdot D_{j}\bm{u}_{j}\,.
\label{eq:MGN_ws}
\end{align} 
In each layer, we define a vertical \emph{material} coordinate, $0\leq s_i \leq 1$, by 
$
s_i=(h_i-z)/D_i
$. Then, by using $d s_i/dt_i+w\partial{s_i}/\partial z=0$ with $d/dt_i = (\partial /\partial  t + \bm{u}_{i}\cdot\nabla)$, we note
\begin{align}
w_{i} = D_i \frac{d s_i}{d t_i}.
\end{align} 

To obtain the new $\sqrt{D}$ equations, we replace $w_{i}$ in (\ref{eq:MGN_lagrangian})
with a different linear approximation of the  vertical motions within
each layer, as follows

\begin{align}
W_{i }& = 
\frac{d_{i}}{D_{i}}\left(\frac{\partial h_{i}}{\partial t} -
\frac{(h_{i}-z)}{D_i} 
\frac{\partial D_{i}}{\partial t} 
\right)
=
d_i\frac{\partial s_i}{\partial t}\:.
\label{eq:W_def}
\end{align}
 This approximation introduces a set of $N$ new length scales, $d_{i}$,
the far field fluid thicknesses. The final quantity in (\ref{eq:W_def}) is a vertical fluid velocity, expressed in the
\emph{convective} representation \citep{HoMaRa1986}. The spatial and convective representations
of fluid dynamics are the analogues, respectively, of the spatial
and body representations of rigid body dynamics on $SO(3)$. This
analogy arises because the configuration spaces for fluid dynamics
and for rigid bodies are both Lie groups. In both cases, spatial velocities
are right-invariant vector fields, while the convective, or body,
velocities are the corresponding left-invariant vector fields.

On taking variations, the final equations arising from Hamilton's
principle for the new Lagrangian may be written compactly in terms
of a shallow water equation, plus additional nonlinear dispersive
terms that represent a non-hydrostatic pressure gradient,\begin{equation}
\frac{\partial\bm{u}_{i}}{\partial t}+\bm{u}_{i}\cdot\nabla\bm{u}_{i}=-g\nabla\hspace{-4mm}\underbrace{\left[h_{i}+\sum_{j=1}^{i-1}\frac{\rho_{j}}{\rho_{i}}D_{j}\right]}_{\hbox{hydrostatic\ pressure}}\hspace{-2mm}-\ \nabla\underbrace{\left[\frac{d_{i}F_{i}}{6}+\sum_{j=1}^{i-1}\frac{\rho_{j}d_{j}G_{j}}{2\rho_{i}}+\frac{d_{i}^{2}E_{i}}{6D_{i}^{2}}\right]}_{\hbox{non-hydrostatic\ pressure}},\label{eq:SRD}\end{equation}
 \[
F_{i}  =\left(\frac{\partial}{\partial t}\frac{d_{i}}{D_{i}}\frac{\partial}{\partial t}\right)\left[2h_{i}+h_{i+1}\right],\quad G_{i}=\left(\frac{\partial}{\partial t}\frac{d_{i}}{D_{i}}\frac{\partial}{\partial t}\right)\left[h_{i}+h_{i+1}\right],\]
\[ E_{i}=\left(\frac{\partial h_{i}}{\partial t}\right)^{2}+\frac{\partial h_{i}}{\partial t}\frac{\partial h_{i+1}}{\partial t}+\left(\frac{\partial h_{i+1}}{\partial t}\right)^{2}.\]
 In comparison, the MGN equations have a similar form, but the non-hydrostatic
forces no longer correspond to the gradient of a pressure. Instead,
the MGN equations take the form \begin{equation}
\frac{\partial\bm{u}_{i}}{\partial t}+\bm{u}_{i}\cdot\nabla\bm{u}_{i}=-g\nabla\left[h_{i}+\sum_{j=1}^{i-1}\frac{\rho_{j}}{\rho_{i}}D_{j}\right]-\frac{1}{D_{i}}\nabla\frac{D_{i}^{2}\widetilde{F}_{i}}{6}-\nabla\sum_{j=1}^{i-1}\frac{\rho_{j}D_{j}\widetilde{G}_{j}}{2\rho_{i}}-\frac{1}{2}\widetilde{G}_{i}\nabla h_{i+1},\label{eq:MGN}\end{equation}
 \[
\widetilde{F_{i}}=\left(\frac{\partial}{\partial t}+\bm{u}_{i}\cdot\nabla\right)^{2}\left[2h_{i}+h_{i+1}\right],\quad\widetilde{G_{i}}=\left(\frac{\partial}{\partial t}+\bm{u}_{i}\cdot\nabla\right)^{2}\left[h_{i}+h_{i+1}\right].\]

 Having arisen from Hamilton's principle, the $\sqrt{D}$ and MGN
equations may both be written in Hamiltonian form. In addition, since
their Lagrangians are both invariant under particle relabelling, the
two sets of equations each possesses the corresponding materially-conserved
quantities,

\[
\frac{\partial q_i}{\partial t}
+
\bm{u}_{i}\cdot\nabla{q}_i 
= 
0, 
\quad\hbox{where}\quad 
q_i := 
\frac{1}{D_{i}}\bm{\hat{z}}\cdot\nabla\times\left(\frac{1}{D_{i}}\frac{\delta\ell}{\delta\bm{u}_{i}}\right) 
. 
\]The quantity $q_{i}$ is the potential vorticity in the $i$-th layer.
For the $\sqrt{D}$ equations, 

\[
q_i=\bm{\hat{z}}\cdot\left(\nabla\times\bm{u}_{i}\right)/D_{i}, 
\] just as for the unmodified shallow water equations. Hence, equation
(\ref{eq:SRD}) admits potential flow solutions for which $\nabla\times\bm{u}_{i}=0$.
In contrast, the MGN potential vorticity contains higher derivatives
of $\bm{u}_{i}$. 

When restricted to a single layer, the Lagrangian for the $\sqrt{D}$
equations (\ref{eq:SRD}) is 
\[
\ell=\frac{\rho}{2}\int D|\bm{u}|^{2}+\frac{4d^{2}}{3}\left(\frac{\partial}{\partial t}\sqrt{D}\right)^{2}-g\left[\left(D-b\right)^{2}-b^{2}\right]\mathrm{d}x\,\mathrm{d}y,
\]
and the single-layer $\sqrt{D}$ motion equation becomes
\[
\frac{\partial\bm{u}}{\partial t}+\bm{u}\cdot\nabla\bm{u}=
-g\nabla\left(D-b\right)-\frac{d^{2}}{6}\nabla\left(\frac{1}{\sqrt{D}}\frac{\partial^{2}\sqrt{D}}{\partial t^{2}}\right).
\]
The single-layer Lagrangian contains a term that coincides with the Fisher-Rao
metric in probability theory. (See \citet{BrHu1998} for an excellent
review of the subject.) This feature suggested the name ``$\sqrt{D}$ equations''
for the new system.

\section{Linear Dispersion Analysis}

This section shows that the $\sqrt{D}$ equations in (\ref{eq:SRD})
possess the same linear dispersion relation at rest as MGN, but that unlike
MGN they remain linearly well-posed when background shear is present.
As discussed in \citet{LB1997}, linear well-posedness requires that
the phase speed remain bounded as $k\to\infty$ for all background
shear profiles. In what follows we specialise to the case $N=2$,
and impose a rigid lid constraint, $h_1 = 0$, through an additional barotropic pressure term, although the analysis generalizes to an arbitrary number of layers and to the free surface case. Linearizing equations (\ref{eq:SRD})
and (\ref{eq:D_continuity}) for far field depths, $d_{1}$, $d_{2}$
and background velocities, $U_{1}$, $U_{2}$ produces the following
dispersion relation

\[
\rho_{1}d_{2}\left(U_{1}-\lambda\right)^{2}+\rho_{2}d_{1}\left(U_{2}-\lambda\right)^{2}+\left({\lambda^{2}k^{2}d_{1}d_{2}}/{3}\right)\left(\rho_{1}d_{1}+\rho_{2}d_{2}\right)-(\rho_{2}-\rho_{1})gd_{1}d_{2}=0.\]
 Here the phase speed is $\lambda=\omega/k$ for frequency $\omega$
and wave number $k$. Meanwhile under the same assumptions the equivalent CC equations
linearize to\[
\rho_{1}d_{2}\left(U_{1}-\lambda\right)^{2}\left(1+{d_{1}k^{2}}/{3}\right)+\rho_{2}d_{1}\left(U_{2}-\lambda\right)^{2}\left(1+{d_{2}k^{2}}/{3}\right)-(\rho_{2}-\rho_{1})gd_{1}d_{2}=0.\]
 When the background state is at rest, $U_{1}=U_{2}=0$, both sets
of equations exhibit the same dispersion relation,\[
\lambda=\frac{\pm\sqrt{\left(\rho_{2}-\rho_{1}\right)gd_{1}d_{2}}}{\sqrt{\rho_{1}d_{2}+\rho_{2}d_{1}+\left(\rho_{1}d_{1}+\rho_{2}d_{2}\right)d_{1}d_{2}k^{2}}},\]
 in which the phase speed of the high wave number modes converges
towards zero. This is not surprising, since when linearized at rest
the two measures of vertical motion coincide in (\ref{eq:MGN_ws})
and (\ref{eq:W_def}), with $w_{i}=W_{i}$.

In the presence of background shear, $U_{1}\ne U_{2}$, the solution
for the phase speed $\lambda$ of the rigid lid $\sqrt{D}$ equations is 
\begin{align*}
\lambda & =\frac{-B_{0}\pm\sqrt{B_{0}^{2}-4\left(A_{0}+A_{2}k^{2}\right)\left(C_{0}-(\rho_{2}-\rho_{1})gd_{1}d_{2}\right)}}{2\left(A_{0}+A_{2}k^{2}\right)},\\
 & \sim\pm\frac{\sqrt{\left(C_{0}-(\rho_{2}-\rho_{1})gd_{1}d_{2}\right)}}{\sqrt{A_{2}}|k|}\qquad\textrm{for }k\gg1
\end{align*}
 and for the CC equations is 
\begin{align*}
\lambda & =\frac{-\left(B_{0}+B_{2}k^{2}\right)\pm\sqrt{\left(B_{0}+B_{2}k^{2}\right)^{2}-4\left(A_{0}+A_{2}k^{2}\right)\left(C_{0}+C_{2}k^{2}-(\rho_{2}-\rho_{1})gd_{1}d_{2}\right)}}{2\left(A_{0}+A_{2}k^{2}\right)},\\
 & \sim\frac{-B_{2}\pm\sqrt{B_{2}^{2}-4A_{2}C_{2}}}{2A_{2}}\qquad\textrm{for }k\gg1.
\end{align*}
 Here the coefficients $A_0$, $B_0$, $C_0$, etc. are functions of the
mean depth ratio, $d_{1}/d_{2}$ and of the shear $(U_{2}-U_{1})$.
\citet{LB1997} showed that the discriminant $B_{2}^{2}-A_{2}C_{2}$
is strictly negative in the CC case, thereby ensuring the existence
of a root with growth rate, $\Im(\lambda k)$, which is positive and
scales linearly with $k$. Thus, the CC equations are linearly ill-posed.
In contrast, the magnitude of growth rates in the $\sqrt{D}$ case
are bounded from above, independently of wave number $k$. Thus, the
$\sqrt{D}$ equations are linearly well-posed. Indeed, provided a
Richardson-number condition is satisfied, that 
\begin{equation}
Ri:= \left(\rho_{2}-\rho_{1}\right)gd_{1}d_{2}\left(d_{1}+d_{2}\right)^{2}
/\Big[\left(\rho_{1}d_{2}^{3}+\rho_{2}d_{1}^{3}\right)|U_{2}-U_{1}|^{2} \Big]
\ge1,
\label{eq:Richardson-condition}
\end{equation}
then the linearized system remains stable to perturbations at \emph{any}
wave number.

\section{Travelling Wave Solutions}

At this point, we have shown that the $\sqrt{D}$ equations satisfy
requirements \emph{(a)} and \emph{(b)} in the Introduction. We will
now show that the $\sqrt{D}$ and MGN equations also satisfy requirement
\emph{(c)}; that is, they share the same travelling wave solutions
for any number of layers, in the absence of imposed background shear. We take a travelling wave ansatz for a choice of wave speed, $c$,
\[
D_{i}(x,t) =: D_{i}(X)
,\quad  
u_{i}(x,t) =:  u_{i}(X) 
,\quad\hbox{with}\quad  
X:=x-ct 
\,.
\] 
Integrating the continuity equation (\ref{eq:D_continuity}) in a
frame moving at the mean barotropic velocity implies for both the
MGN and $\sqrt{D}$ systems that \begin{equation}
u_{i}=c\left(1-{d_{i}}/{D_{i}}\right).\label{eq:u_travelling_wave}\end{equation}
 %
{}For such solutions we see from equations (\ref{eq:MGN_ws}) and (\ref{eq:W_def}) that $w_{i}$ and $W_{i}$ coincide, \begin{equation}
w_{i}
=
(h_{i}-z)\frac{\partial u_{i}}{\partial X}+(u_{i}-c)\frac{\partial h_{i}}{\partial X}
=
-\frac{cd_{i}}{D_i}\left[\frac{(z-h_{i})}{D_{i}}\frac{\partial D_{i}}{\partial X}+\frac{\partial h_{i}}{\partial X}\right]
=W_{i}.\label{eq:TW_w_equivalence}
\end{equation}
 To obtain travelling wave solutions, one varies the Lagrangian subject to the relation (\ref{eq:u_travelling_wave}). Using relation (\ref{eq:TW_w_equivalence}) between the vertical velocities implies that the travelling
wave Lagrangians for the $\sqrt{D}$ and MGN equations coincide. Consequently,
the travelling wave solutions of the two equation sets must also coincide. This finishes the derivation of the new equation set and the verification of the three
requirements set out in the Introduction. In the presence of background shear, however, $(w_i-W_i)\ne0$, and the coincidence of the travelling waves of the two models no longer holds. 

\section{Numerical experiments (Single layer)}

We now present some results of numerical computation with the $\sqrt{D}$
equations for a single layer. The equations are discretized using
standard finite volume techniques, without any filtering. In this
section we compare results for the one-layer $\sqrt{D}$ and GN equations.
These results show that the two systems have the same qualitative
behaviour.

\subsection{Single layer lock release\label{sub:lock_release}}

\begin{figure}
\begin{centering}
\includegraphics[width=1\textwidth]{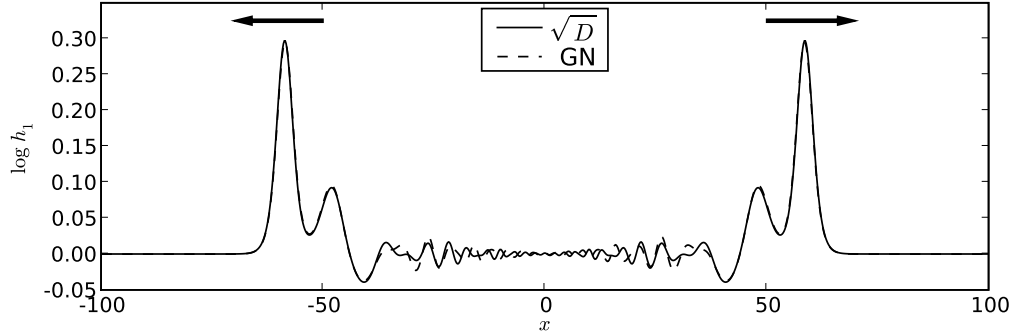}
\par\end{centering}

\caption{Single-layer $\sqrt{D}$ and GN thickness profiles arising from lock-release runs. The arrows indicate directions of crest propagation. The leading wave profiles are essentially identical for both the $\sqrt{D}$ and GN equations. The smaller crests are also very similar. (Differences are emphasised by using a logarithmic scale.)}

\label{Flo:figure_lock-release} 
\end{figure}

We first consider a lock release experiment with periodic boundary
conditions and physical parameters and initial conditions given by\[
d=1,\quad g=1,\]
 \[
D(x,0)=1+0.5\left[\tanh\left(x-4\right)-\tanh\left(x+4\right)\right],\quad u(x,0)=0,\quad x\in[-100,100].\]
 We integrated this state forwards in time for both the GN and
$\sqrt{D}$ systems. Figure \ref{Flo:figure_lock-release} shows a snapshot
of a train of travelling waves emerging from the lock release in each equation set. In
both cases, the first wave is the largest and the wave profiles of
the two systems track each other closely. This may have been expected,
because the two systems share the same travelling wave solutions and
possess identical linear dispersion relations for a quiescent background.\rem{
\begin{framed}
Needs to be much stronger here. 
\end{framed}
The profile of the leading order response is also very close to the true travelling waves of the later experiments. Since the wave speed of such travelling waves is proportional to the root of the maximum amplitude, this means that the leading order signal in the far field will also be identical. Plotting the profiles for comparison shows that after sufficient time the subsequent signal is also closely matched, with some discrepancy in phase, but little in maximum amplitude.}

\subsection{Single layer interaction of solitary waves}

\begin{figure}
\begin{centering}
\includegraphics[width=1\textwidth]{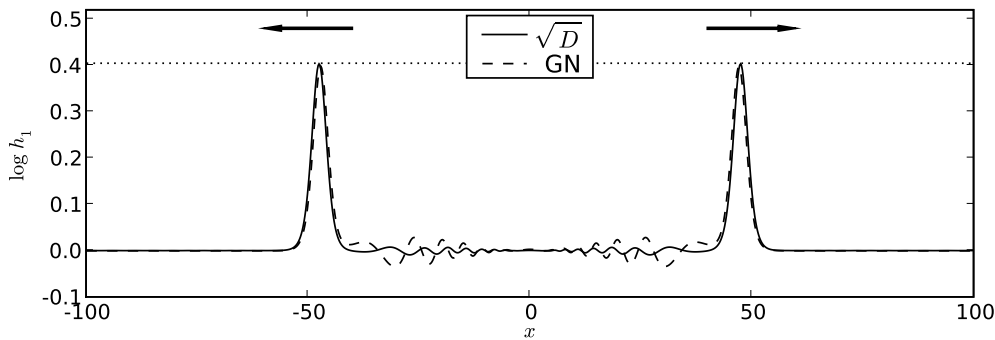}
\par\end{centering}

\caption{Single-layer $\sqrt{D}$ and GN results are shown for symmetric head-on
collision experiments. Layer thickness profiles are shown after the collision. The fully nonlinear collision is near elastic
for both $\sqrt{D}$ and GN, with near conservation of the initial
amplitude, shown by the dotted line. The travelling wave behaviour
is essentially the same, except for a very small phase shift, while
the slow, small-amplitude radiation waves show a few minor differences.}

\label{Flo:figure_1layer_collisions} 
\end{figure}

We now consider the interaction between pairs of solitary travelling
wave solutions in the two systems. The $\sqrt{D}$ and GN systems
have the same travelling wave solutions, but their PDEs differ; so
one may expect to see differences in their wave-interaction properties.
For both systems the travelling wave takes a $\textrm{sech}^{2}$
form,\[
D=d\left[1+\left(\frac{c^{2}}{gd}-1\right)\textrm{sech}^{2}\left(\frac{\sqrt{3\left(c^2-gd\right)}\left(x-ct\right)}{2cd}\right)\right],\]
 and the velocity profile is given by equation (\ref{eq:u_travelling_wave}).
Figure \ref{Flo:figure_1layer_collisions} shows a snapshot of the results for the case
of two equal amplitude, oppositely directed travelling waves after suffering
a symmetric head-on collision. Both the $\sqrt{D}$ and GN descriptions
show a near-elastic collision followed by small-amplitude radiation.
\rem{\foreignlanguage{british}{This results in a slight loss in the wave amplitude through the impact, as well as a small phase lag. However, the magnitude of this lag is slightly different in the two systems with the new  equations showing a smaller phase shift (and thus a smaller interaction time).}}
The travelling wave behaviour is essentially identical. The main difference
lies in the details of the slower linear waves that are radiated and
left behind during the nonlinear wave interactions. Similar results
were obtained in the case of asymmetric collisions.

\section{Numerical Experiments (Two Layers, Rigid Lid)}

We now consider numerical solutions of the two-layer rigid lid $\sqrt{D}$
equations in one spatial dimension. Unlike the result of \citet{JoChoi2002}
for the CC equations, there is no value of resolution at which the
code for numerically integrating the $\sqrt{D}$ equations becomes
unstable. This might have been hoped, because the $\sqrt{D}$ equations
are linearly well-posed.

\subsection{Two layer rigid lid lock release experiments}

\begin{figure}
\centering{}\includegraphics[width=1\textwidth]{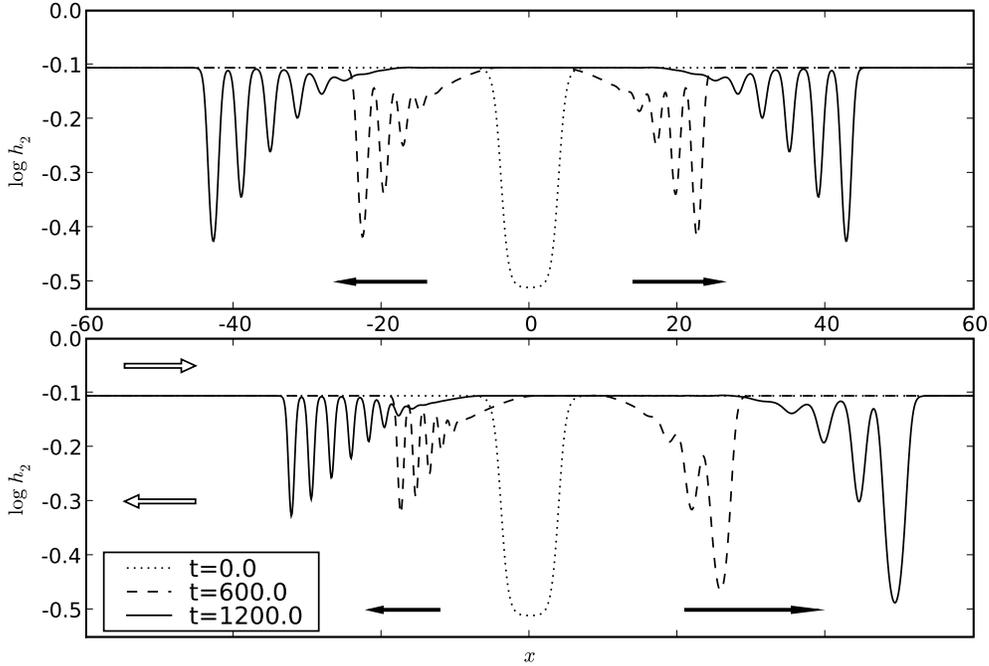}
\caption{Two-layer rigid lid lock release experiments are compared, with and
without background shear of Richardson number $Ri=5.02$. The fluid interface is plotted at times $t=0$, 600, 1200. Here $g=1$, $\rho_{1}/\rho_{2}=0.995$,
$d_{1}/h_{1}=0.1$. The fluid was initialised with a symmetric
tanh profile in thickness. Filled arrows denote the directions of travel
of the waves, while unfilled arrows indicate the background shear.
The run without shear (upper figure) generated a left-right symmetric distribution of wave crests. In contrast, the run with background shear (lower figure) generated
an asymmetric distribution of wave peaks. The lower figure shows lock release for an upper layer moving with initially constant rightward velocity. Consequently, there was a net transfer of momentum toward the right, which significantly affected the distribution of wave crests. 
\label{Flo:figure_two_layer_lock-release} }

\end{figure}

We next present the result of the equivalent lock-release experiment
to that of Section \ref{sub:lock_release} for the two-layer rigid lid
equations in lock release configuration. For this, we choose $d_{1}=0.1$, $d_{2}=0.9$, $g=1$, and
 \[
D_{1}(x,0)=d_{1}+0.3\left[\tanh\left(x-4\right)-\tanh\left(x+4\right)\right],\;
D_{2}=d_{1}+d_{2}-D_{1},\; x\in[-100,100].\]
 The initial configuration is taken as a depression of the fluid interface,
because CC or $\sqrt{D}$ travelling waves extend from the thinner layer
into the thicker one. Since our motivating interest was in modelling
waves on the ocean thermocline, we will concentrate here on the situation
in which the top layer is thinner than the bottom layer. We show results
for the following velocity profiles:
\begin{enumerate}
\item No shear:
 $u_{1}(x,0)=0$, $u_{2}(x,0)=0$, and $x\in[-100,100]$.

\item Background shear flow: $u_{1}(x,0)=0.01$,
  $u_{2}=-{D_{1}u_{1}}/{D_{2}}$,amd  $x\in[-100,100].$

\end{enumerate}
From (\ref{eq:Richardson-condition}) the Richardson number for the case with shear is $Ri=5.02$, and the flow is still linearly stable for all wave numbers. 

\begin{figure}
\centering{}\includegraphics[width=1\textwidth]{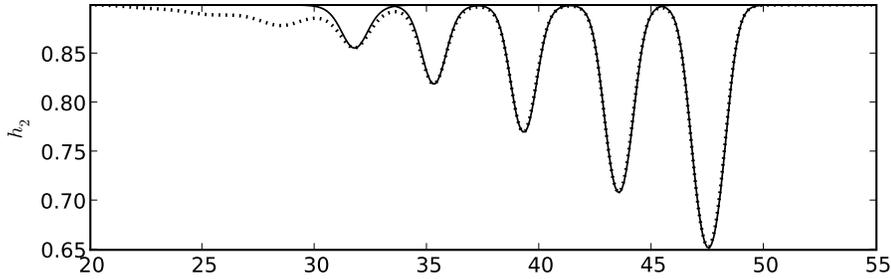}\caption{Close up of two-layer wave lock-release experiment in Figure \ref{Flo:figure_two_layer_lock-release}. The corresponding travelling
wave profile (solid line, as defined by the maximum wave amplitude)
is plotted over the numerical lock-release solution at t=1300 (dashed
line). In the three leading waves, the differences from the corresponding
solitary wave profiles are indiscernible.
\label{Flo:closeup}}

\end{figure}

The result of the integration is shown in Figure \ref{Flo:figure_two_layer_lock-release}.
Unlike the single-layer case, the linear long wave speed in the two-layer case is close to the wave speed of the resulting travelling
waves. In Figure \ref{Flo:closeup} we compare the leading short wave
disturbances of one lobe in the shear-free lock-release experiment to
the calculated solitary travelling wave solutions in the same parameter
regime, as defined by the maximum amplitude of the disturbance. For
the three largest waves the differences are indistinguishable, while the
smaller waves, which are still interacting with the long wave signal
are wider than the equivalent true travelling wave. A similar
result holds for the asymmetric crests generated in the presence of
shear and for the respective velocity profiles. The leading order
solution is thus well approximated as a train
of independent solitary travelling waves propagating in order of amplitude.

\subsection{Interaction of two layer solitary waves with rigid lid}

\begin{figure}
\begin{centering}
\includegraphics[width=1\textwidth]{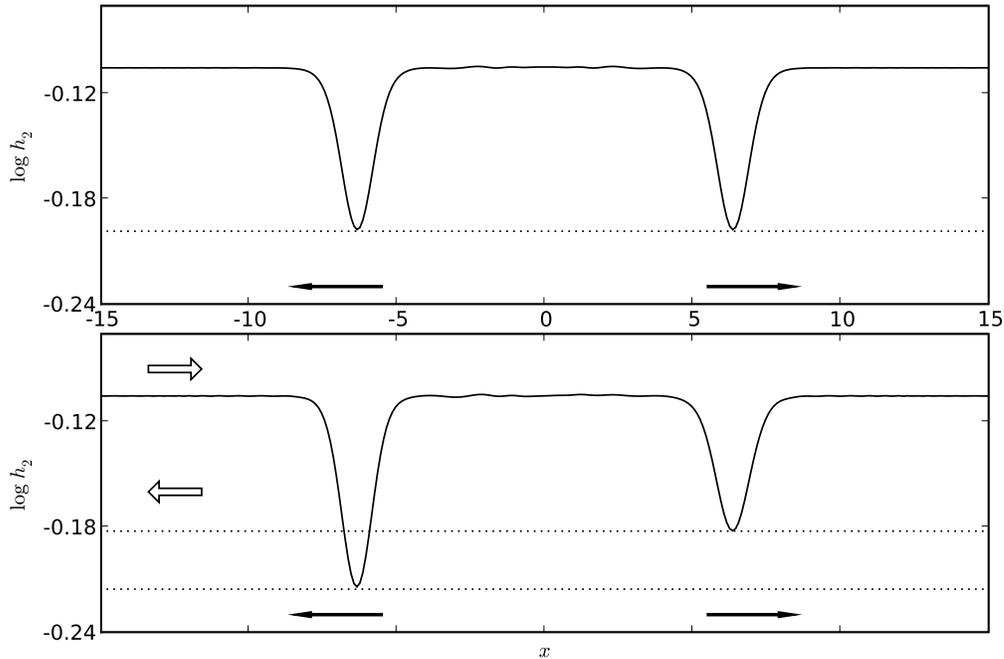}\caption{The figure shows two-layer head-on collision experiments with a rigid lid.
In the top figure there is no background shear, while in the bottom figure the Richardson number is $Ri=502$. The interface depth is shown after the symmetric
head-on collision of two travelling waves of the same phase speed.
Physical parameters are, $g=1$, $\rho_{1}/\rho_{2}=0.995$, $d_{1}/h_{1}=0.1$.
In both figures the phase velocity is $c/\sqrt{gh_1}=\pm0.027$ and the initial amplitudes are indicated by dotted lines. Filled arrows denote the directions in which the solitary waves are  travelling, while unfilled arrows indicate the background shear. In both cases, the collision was found to be nearly elastic, and the final conditions were essentially mirror reflections of the initial conditions, with only slight changes in the amplitudes. 
\label{Flo:figure_two_layer_collisions}
}

\par\end{centering}

\end{figure}

Finally we consider wave interaction experiments in the two-layer
case. Here we initialize with two numerical solutions of the
travelling wave problem with oppositely directed velocities. Results
for these two experiments are similar to those in the single layer
case and also similar to those in \citet{JoChoi2002} in the case
without shear.  Figure \ref{Flo:figure_two_layer_collisions} shows our
numerical solution following the wave collision in the case of equal
magnitude wave speed with and without a background shear. We again see
nearly elastic collisions, after which the waves approximately regain
both their shape and amplitude. In the case with shear, the wave
profiles are no longer symmetric.  Defining upstream and downstream
direction in terms of the background flow in the thicker layer, we see
that the upstream gains amplitude and narrows compared with the
shear-free wave profile at the same wave speed, while the wave
travelling upstream loses amplitude and widens. Nevertheless, the
waves are seen to nearly conserve amplitude, and hence momentum
through the collision.

%

\section{Summary}

We have introduced a system of Hamiltonian multilayer water wave equations
that is linearly well-posed in the presence of background shear but
possesses the same travelling wave solutions as the MGN equations,
whose ill-posedness had previously caused difficulties in their numerical integration. The new system also has been found numerically
to generate fast-moving trains of large-amplitude coherent waves that
exhibit ballistic, nearly-elastic, nonlinear scattering behaviour
amidst a background of slow, small-amplitude, weakly interacting or
linear wave radiation. Future work will explore wave-wave and wave-topography
interactions in the presence of vertical shear between the
layers, as well as two-dimensional multilayer wave interactions and
wave generation by flow over topography by using the new well-posed
$\sqrt{D}$ equation set.

We are grateful to W. Choi, F. Dias and R. Grimshaw for fruitful discussions
on this topic. J. R. Percival was supported by the ONR grant N00014-05-1-0703.

\bibliographystyle{jfm2}
\bibliography{MLCM}

\begin{thebibliography}{11}
\expandafter\ifx\csname natexlab\endcsname\relax\def\natexlab#1{#1}\fi

\bibitem[Barros \& Choi(2009)]{barros2009isi}
{\sc Barros, R. \& Choi, W.} 2009 Inhibiting shear instability induced by large
  amplitude internal solitary waves in two-layer flows with a free surface.
  {\em Studies in Applied Mathematics\/} {\bf 122}~(3), 325--346.

\bibitem[Brody \& Hughston(1998)]{BrHu1998}
{\sc Brody, D.~C. \& Hughston, L.~P.} 1998 Statistical geometry in quantum
  mechanics. {\em Proc. Roy. Soc. A\/} {\bf 454}, 2445--2475.

\bibitem[Choi \& Camassa(1996)]{CC96JFM}
{\sc Choi, W. \& Camassa, R.} 1996 Weakly nonlinear internal waves in a
  two-fluid system. {\em Journal of Fluid Mechanics\/} {\bf 313}, 83--103.

\bibitem[Choi \& Camassa(1999)]{CC99}
{\sc Choi, W. \& Camassa, R.} 1999 Fully nonlinear internal waves in a
  two-fluid system. {\em Journal of Fluid Mechanics\/} {\bf 396}, 1--36.

\bibitem[Green \& Naghdi(1976)]{GN76.}
{\sc Green, A.~E. \& Naghdi, P.~M.} 1976 A derivation for wave propagation in
  water of variable depth. {\em Journal of Fluid Mechanics\/} {\bf 78},
  237--246.

\bibitem[Holm {\em et~al.\/}(1986)Holm, Marsden \& Ratiu]{HoMaRa1986}
{\sc Holm, D.~D., Marsden, J.~E. \& Ratiu, T.~S.} 1986 The {H}amiltonian
  structure of continuum mechanics in material, inverse material, spatial and
  convective representations. In {\em {Hamiltonian Structure and Lyapunov
  Stability for Ideal Continuum Dynamics}\/}. University of Montreal Press,.

\bibitem[Holm {\em et~al.\/}(1998)Holm, Marsden \& Ratiu]{HMR1998}
{\sc Holm, D.~D., Marsden, J.~E. \& Ratiu, T.~S.} 1998 The
  {E}uler-{P}oincar\'{e} equations and semi-direct products with applications
  to continuum therories. {\em Advances in Mathematics\/} .

\bibitem[Jo \& Choi(2002)]{JoChoi2002}
{\sc Jo, T.-C. \& Choi, W.} 2002 Dynamics of strongly nonlinear internal
  solitary waves in shallow water. {\em Studies in Applied Mathematics\/} .

\bibitem[Liska \& Wendroff(1997)]{LB1997}
{\sc Liska, R. \& Wendroff, B.} 1997 Analysis and computation wth stratified
  fluid models. {\em Journal of Computational Physics\/} {\bf 137}~(1).

\bibitem[Liu {\em et~al.\/}(1998)Liu, Chang, Hsu \& Liang]{LCHL98}
{\sc Liu, A.~K., Chang, Y.~S., Hsu, M.~K. \& Liang, N.~K.} 1998 Evolution of
  nonlinear internal waves in the {E}ast and {S}outh {C}hina {S}eas. {\em
  Journal of Geophysical Research\/} .

\bibitem[Percival {\em et~al.\/}(2008)Percival, Holm \& Cotter]{PCH2008}
{\sc Percival, J.~R., Holm, D.~D. \& Cotter, C.~J.} 2008 A
  {E}uler-{P}oincar{\'e} framework for the {G}reen-{N}aghdi equations. {\em
  Journal of Physics A: Mathematical and Theoretical\/} {\bf 41}~(34).

\end{thebibliography}

\end{document}